\documentclass[aps, pra, superscriptaddress, twocolumn, amsfonts, amsmath, amssymb, floatfix]{revtex4-2}
\usepackage{graphicx}
\usepackage{float}
\usepackage{sidecap}
\usepackage{dcolumn}
\usepackage{bm}
\usepackage{epsfig}
\usepackage{braket}
\usepackage{color}
\usepackage{ulem}
\usepackage{amsmath}
\usepackage{wasysym}
\usepackage[colorlinks=true, letterpaper=true, pdfstartview=FitV, linkcolor=blue, citecolor=blue, urlcolor=blue]{hyperref}

\begin{document}

\title{Collective Oscillations of Bose-Einstein Condensates in a Synthetic Magnetic Field}

\author{Huaxin He}
\thanks{These authors contributed equally to this work.}
\affiliation{Institute for Quantum Science and Technology, Department of Physics, Shanghai University, Shanghai 200444, China}

\author{Fengtao Pang}
\thanks{These authors contributed equally to this work.}
\affiliation{Institute for Quantum Science and Technology, Department of Physics, Shanghai University, Shanghai 200444, China}

\author{Yongping Zhang}
\email{yongping11@t.shu.edu.cn}
\affiliation{Institute for Quantum Science and Technology, Department of Physics, Shanghai University, Shanghai 200444, China}

\author{Chunlei Qu}
\email{cqu5@stevens.edu}
\affiliation{Department of Physics, Stevens Institute of Technology, Hoboken, NJ 07030, USA}
\affiliation{Center for Quantum Science and Engineering, Stevens Institute of Technology, Hoboken, NJ 07030, USA}

\begin{abstract}
	We study the collective oscillations of spin-orbit-coupled Bose-Einstein condensates in the presence of position-dependent detuning. Specifically, we explore the quadrupole modes of the system using both numerical and analytical approaches based on the Gross-Pitaevskii equation and hydrodynamic theory. Due to spin-orbit coupling and the synthetic magnetic field, {the $xy$ scissors mode couples with a superposition of the three diagonal quadrupole modes ($x^2$, $y^2$, and $z^2$),} resulting in the characteristic beating effect. {The remaining two scissors modes, $xz$ and $yz$, are coupled, giving rise to a Lissajous-like pattern that is highly sensitive to the excitation method and orientation of the synthetic magnetic field.} Furthermore, we find that anisotropic interactions as well as the direction of the synthetic magnetic field, can significantly influence the oscillation amplitude and frequency of the quadrupole modes. These findings highlight the potential of Bose-Einstein condensates under synthetic magnetic fields for quantum sensing applications, such as magnetic field {gradient} measurements, and provide a promising foundation for future experimental research and technological development.
\end{abstract}

\maketitle

\section{Introduction}
Ultracold atomic gases offer unparalleled controllability and flexibility, making them an ideal platform for quantum simulations to explore exotic states of matter~\cite{anderson1995observation,o2002observation,lewenstein2007ultracold,RevModPhys.80.885}. Artificial gauge fields enable neutral atoms to mimic electrons in sensing electromagnetic fields, facilitating the investigation of synthetic magnetism and synthetic spin-orbit coupling (SOC) using ultracold atoms~\cite{PhysRevLett.102.130401,lin2009synthetic,lin2011synthetic,lin2011spin,goldman2014light,huang2016experimental,wu2016realization}. It is well known that magnetic fields and SOC are key origins of topological physics in solid-state electronic systems~\cite{RevModPhys.82.3045}. Ultracold atoms, when coupled with artificial gauge fields, offer a novel means to explore and substantiate various topological phenomena~\cite{RevModPhys.83.1523}. Notably, topological insulator models such as the Haldane model and SSH model have been successfully implemented in recent ultracold atom experiments, enabling the study of their topological properties with neutral atoms~\cite{jotzu2014experimental,atala2013direct,lohse2016thouless,leder2016real,RevModPhys.91.015005,PhysRevLett.112.086401}.

{The rapid progress in SOC Bose-Einstein condensates (BEC) has stimulated extensive research into their exotic superfluid properties~\cite{zhu2012exotic,PhysRevA.86.063621,PhysRevA.85.053607,PhysRevA.86.041604,PhysRevLett.109.115301,PhysRevA.87.063610,qu2017spin1,PhysRevA.84.063604,PhysRevA.89.023629,PhysRevA.95.043605,PhysRevA.84.063624,PhysRevA.95.033603,PhysRevA.89.061605,PhysRevLett.105.160403,PhysRevLett.107.150403,PhysRevLett.108.225301,PhysRevLett.108.010402,PhysRevA.94.033635,PhysRevLett.118.145302,PhysRevLett.120.183202,PhysRevA.108.053316,zheng2013properties,PhysRevLett.114.070401,li2017stripe}. Compared to traditional BEC, SOC introduces a spin-dependent modification in the momentum operator, which fundamentally modifies the current-phase relationship~\cite{PhysRevA.87.063610,qu2017spin1}. This modification leads to the violation of the rotational constraint of the velocity field~\cite{PhysRevLett.118.145302}, which arises due to the breaking of Galilean invariance induced by SOC~\cite{zheng2013properties,PhysRevLett.114.070401}. A direct consequence of this symmetry breaking is that the definition of the superfluid critical velocity depends on the observer's reference frame~\cite{zhu2012exotic}, while the velocity field associated with the fluid rotation exhibits diffusive vorticity~\cite{PhysRevLett.118.145302}.  Beyond these fundamental changes in superfluidity, SOC BEC also exhibit novel quantum phases and dynamics. For instance, when the Raman coupling is weak, the atoms may occupy two momentum states, leading to the emergence of a stripe phase—a novel state that exhibits both crystalline-like order and superfluid behavior~\cite{PhysRevLett.105.160403,PhysRevLett.108.010402,li2017stripe}. When Raman coupling is increased, the condensate may evolve from a plane-wave phase to a zero-momentum phase, where many interesting physical quantities diverge~\cite{PhysRevA.86.063621,PhysRevA.94.033635}. Furthermore, introducing a position-dependent detuning into the SOC BEC system creates a synthetic magnetic field for neutral atoms~\cite{PhysRevA.79.063613}, which imparts nonzero angular momentum and leads to a rigid-like rotational velocity field~\cite{PhysRevLett.120.183202}. As the detuning gradient increases, vortices can be generated without the rotation of the confining trap~\cite{PhysRevA.108.053316}, demonstrating a novel mechanism for vortex formation. These findings collectively highlight the rich interplay between SOC and superfluidity and open new avenues for exploring quantum many-body physics.}

Collective modes play a crucial role in revealing fundamental properties of ultracold atomic gases, such as testing superfluidity, calibrating the frequency of trapping potentials, and detecting angular momentum~\cite{PhysRevLett.77.2360,RevModPhys.71.463,PhysRevLett.83.4452,PhysRevLett.84.2056,PhysRevA.69.043621}. The dynamics of SOC BEC in a synthetic magnetic field exhibit novel properties, including the precession of dipole oscillations, similar to a Foucault pendulum, and a Hall-like effect produced by quench dynamics~\cite{doi:10.1073/pnas.1202579109,PhysRevLett.120.183202}. The synthetic magnetic field induces the coupling between the scissors and quadrupole modes leading to a notable beating effect and gyroscope-like dynamics~\cite{PhysRevA.108.053316}. Moreover, recent studies have uncovered an interesting spin-deflection effect by the rigid-like rotational velocity field during the expansion dynamics~\cite{PhysRevA.110.043307}. In exploring these collective dynamics, {the recent work~\cite{PhysRevA.108.053316}} has primarily focused on a specific parameter regime where the quadrupole modes in the plane perpendicular to the synthetic magnetic field are degenerate. Under the influence of synthetic magnetic fields, these degenerate modes exhibit strong coupling, leading to pronounced collective dynamics.

In this paper, we systematically investigate collective oscillations of SOC BEC in a synthetic magnetic field by considering a broad parameter regime. Our results generalize the application of spinor hydrodynamic (HD) theory and are corroborated by the numerical simulation of the Gross-Pitaevskii equation (GPE). We find that various quadrupole modes are coupled with the scissors mode, resulting in a perfect beating effect in their collective oscillations. Additionally, the oscillation trajectories of {the other two} coupled scissors modes exhibit a feature similar to Lissajous patterns. {Additionally, the oscillation trajectories of the other two coupled scissors modes resemble Lissajous patterns—trajectory curves generated by the combination of two harmonic vibrations that are perpendicular to each other~\cite{PhysRevE.91.062901,PhysRevLett.123.214101}.} We have also extended our calculation to the system with anisotropic interactions between the two components, which are crucial for a general understanding of the real system. Additionally, we study the collective oscillation of the BEC in a synthetic magnetic field that points to an arbitrary direction. Our findings reveal that anisotropic interactions and the direction of synthetic magnetic fields significantly alter the oscillation amplitude, frequency, and coupling of these modes.

This paper is organized as follows. In Sec.~\ref{sec2}, we describe the basic model of the system and the derivation of the HD equations. In Sec.~\ref{sec3}, we study the coupling dynamics of quadrupole modes by comparing numerical simulations and analytical results from HD. In Sec.~\ref{sec4} and \ref{sec5}, the model is extended to more general cases, considering the effects of anisotropic interactions and different directions of synthetic magnetic fields on quadrupole mode coupling. Finally, Section~\ref{sec6} provides a summary of the paper.

\section{Model}\label{sec2}

We consider a spin-1/2 BEC of $^{87}$Rb atoms at zero temperature, with an equal Rashba-Dresselhaus SOC induced along the $x$-direction by two-photon Raman coupling~\cite{lin2011spin,PhysRevLett.108.035302,goldman2014light,zhai2015degenerate,zhang2016properties}. In the mean-field framework, the dynamics of the SOC BEC are governed by the {(hereafter set $\hbar  = 1$)}
\begin{equation}
i\partial_t \psi = (H_0 +H_{\text{int}} ) \psi,
\label{GPE}
\end{equation}
where $\psi=(\psi_1,\psi_2)^T$ represents the order parameter of the two components, satisfying the normalization condition $\int d\textbf{r}|\psi|^2 = N$, with $N$ denoting the total number of atoms. The two-body interactions between atoms are described by the nonlinear term $H_{\text{int}} = \text{diag}(g_{11}|\psi_1|^2+g_{12}|\psi_2|^2, g_{12}|\psi_1|^2+g_{22}|\psi_2|^2)$, where $g_{ij}=4\pi a_{ij}/m$ ($i,j \in \{1, 2\}$) denotes the intra-species interaction strengths ($g_{11}$, $g_{22}$) or the inter-species interaction strength ($g_{12}$), with $a_{ij}$ representing the corresponding $s$-wave scattering lengths, and $m$ being the atomic mass. For simplicity, we assume that the intra-species interaction strengths are equal, i.e. $ g_{11} = g_{22} = g$. The inter-species interaction strength is characterized by the dimensionless interaction parameter $G = g_{12}/g$. Furthermore, $H_0$ is the single-particle SOC Hamiltonian,
\begin{equation}
H_0=\frac{1}{2m}(\hat{\boldsymbol{p}}- k_0\sigma_z \hat{\boldsymbol{e}}_x)^2+V_{\rm{trap}}-\frac{\Omega}{2}\sigma_x-\eta k_0 y \sigma_z,
\label{eq2}
\end{equation}
where $ k_0$ is the momentum transferred during the two-photon Raman process, $\sigma_{x,y,z}$ are the usual $2 \times 2$ Pauli matrices,  $V_{\text{trap}}=\frac{m}{2} (\omega_x^2x^2+\omega_y^2y^2+\omega_z^2z^2)$ is the 3D harmonic potential with trapping frequencies of $\omega_{x, y, z}$, and $\Omega$ represents the Raman coupling strength. {Additionally, $\eta$ represents the gradient of the position-dependent detuning, given by $\delta(y) = -\eta k_0 y$. In general, when a position-dependent detuning $\delta(\mathbf{r})$ is present, the lower band of the SOC BEC Hamiltonian can be approximated as:  
	\begin{equation}  
	H \simeq \frac{\left(\mathbf{p} - A_x^*(\bm{r})\hat{e}_x\right)^2}{2m},
	\end{equation}  
	where $A_x^*(\bm{r}) = k_{\text{min}}(\delta(\mathbf{r}))$ acts as a synthetic vector potential. The synthetic magnetic field, defined as the curl of the synthetic vector potential, is expressed as: 
	\begin{equation}  
	\textbf{B}_{\text{syn}} = \partial_z A_x^* \, \hat{\bm{e}}_y - \partial_y A_x^* \, \hat{\bm{e}}_z.
	\label{SynB}
	\end{equation}  
	When the detuning depends only on $y$, as discussed in Sec.III and Sec. IV, the resulting synthetic magnetic field $\textbf{B}_{\text{syn}}$ points in the $z$-direction. This field generates a rotational velocity field in the $x$-$y$ plane, imparting nonzero angular momentum to the BEC. Conversely, when the detuning depends on both $y$ and $z$, as discussed in Sec. V, the resulting synthetic field can point to any arbitrary direction in the $y$-$z$ plane.} When $\eta$ exceeds a critical value $\eta_c$, vortices form in both spin components of the BEC. When $\eta$ is sufficiently large, the two spin components of the BEC become separated~\cite{PhysRevA.108.053316}.

The HD theory of superfluids offers a useful perspective for studying the equilibrium configurations and low-lying collective modes of BEC~\cite{pitaevskii2003bose,PhysRevA.86.063621}. Starting from Eq.~(\ref{GPE}), we express the order parameter as $\psi = (\sqrt{n_1} e^{i\phi_1}, \sqrt{n_2} e^{i\phi_2})^T$, where $n_1$, $\phi_1$ and $n_2$, $\phi_2$ represent the density and phase for the two components, respectively. The HD formalism can be developed by deriving the differential equations for these new variables and performing the linearization. 
{Due to the presence of SOC, the superfluid velocity of the condensate is fundamentally altered to $\boldsymbol{v}_0 = (\bm{\nabla} \phi_0 - k_0 s_z \hat{\bm{e}}_x/n_0)/m$, where $\phi_0=(\phi_1+\phi_2)/2$ represents the common phase of the two order parameters, $s_z=n_1-n_2$ denotes the spin density, and $n_0=n_1+n_2$ is the total density.}
To simplify our analytical results, we focus on the zero-momentum phase characterized by $\Omega > \Omega_c$ with isotropic interactions ($G = 1$). Here, $\Omega_c=4E_\text{r}$ is the critical value of the Raman coupling that marks the transition from the zero-momentum phase to the plane-wave phase and $E_\text{r}= k_0^2/(2m)$ denotes the recoil energy. When the detuning coefficient $\eta$ is small, the Thomas-Fermi (TF) distribution can be used to approximate the total density. Thus, at equilibrium, the total density is given by $ n_0 = (\mu - V_{\text{trap}})/g $, where $\mu$ is the chemical potential. The phase $\phi_0$ and spin density $s_z$ are determined as 
\begin{equation}
\phi_0= \alpha xy,\quad s_z = 2 \beta yn,
\end{equation}
with $\alpha$ and $\beta$ given by $ \alpha = (2\eta k_0^2 \omega_x^2)/(\Omega \omega_{xy}^2)$ and $ \beta = \eta k_0(\omega_x^2 + \omega_y^2)/(\Omega \omega_{xy}^2)$~\cite{PhysRevLett.120.183202}. Here, $\omega_{xy} = \sqrt{(m/m^*)\omega_x^2 + \omega_y^2}$ represents the oscillation frequency of the scissors mode $xy$ in the absence of the synthetic magnetic field. {Furthermore, the superfluid velocity field $\boldsymbol{v}_0$ exhibits a rigid-body rotation form: $\boldsymbol{v}_0 = (-\omega_{\text{eff}} y, \omega_{\text{eff}}' x, 0)$. Here, $\omega_{\text{eff}}$ and $\omega_{\text{eff}}^\prime$ are effective frequencies characterizing the system's rotational behavior, with $\omega_{\text{eff}} = \eta \Omega_c / \Omega - \alpha / m^*$, $\omega_{\text{eff}}^\prime = \alpha / m$, and $m^* = m \left(1 - \Omega_c / \Omega \right)^{-1}$ representing the effective mass in the zero-momentum phase.} This rigid-like rotational velocity field induces coupling between various collective modes~\cite{PhysRevLett.120.183202,PhysRevA.108.053316}, resulting in intriguing dynamical phenomena. 

To study collective modes, one can introduce variations in density and phase around their equilibrium configurations, represented as 
$n = n_0 + \delta n$ and $\phi = \phi_0 + \delta \phi$, to derive the corresponding oscillation equations. We assume that the spatial variations in density are smooth not only at equilibrium but also in low-excited states, thereby allowing the quantum pressure term to be effectively neglected. The linearized HD equations for fluctuations in density and phase can be derived as~\cite{PhysRevLett.120.183202,PhysRevA.108.053316}:
\begin{equation}
\begin{aligned}
&\frac{\partial \delta n}{\partial t}+\frac{1}{m^{*}} \nabla_{x}\left[n_{0} \nabla_{x}(\delta \phi)\right]+\frac{1}{m} \nabla_{y}\left[n_{0} \nabla_{y}(\delta \phi)\right] \\
+&\frac{1}{m} \nabla_{z}\left[n_{0} \nabla_{z}(\delta \phi)\right]-\omega_{\text {eff}} y \nabla_{x}(\delta n)+\omega_{\mathrm{eff}}^{\prime} x \nabla_{y}(\delta n)=0, \\
&\frac{\partial \delta \phi}{\partial t}+g \delta n-\omega_{\mathrm{eff}} y \nabla_{x} \delta \phi+\omega_{\mathrm{eff}}^{\prime} x \nabla_{y} \delta \phi=0.
\end{aligned}
\label{raodong}
\end{equation}
By setting various ansatz forms for fluctuations in density and phase, different mode excitations can be studied, which makes the analysis more intuitive. In the following sections, we focus on the coupling between the quadrupole modes.

\section{Quadrupole Modes}\label{sec3}

Oscillatory behavior related to the deformation and rotation of an atomic cloud is connected with the superfluidity of BEC. The phenomenon is effectively described by quadrupole modes, which can be categorized into two types~\cite{PhysRevLett.83.4452,PhysRevLett.84.2056,PhysRevA.69.043621}: the diagonal quadrupole modes, represented by the operators $x^2$, $y^2$ and $z^2$, corresponding to oscillations in the overall shape of the condensate; and the scissors modes, represented by the operators $xy$, $xz$ and $yz$, corresponding to angular rotations of the condensate within a plane. For the SOC system, these modes have been used to detect the Hall effect of superfluid in the experimental system~\cite{doi:10.1073/pnas.1202579109}. The introduction of the synthetic magnetic field leads to the coupling of these six modes. Previous research has primarily concentrated on a specific parameter regime where the dipole modes in the plane perpendicular to the synthetic magnetic field are degenerate~\cite{PhysRevLett.120.183202}. This article aims to conduct a more systematic study on this phenomenon. 

In the framework of HD theory, to investigate the quadrupole modes, we express the density and relative phase fluctuations as quadratic polynomials, which enables us to obtain analytical solutions to Eq.~(\ref{raodong}). The ansatz for these fluctuations can be written as:
\begin{equation}
\begin{aligned}
\delta n=&\epsilon_1\frac{xy}{R_xR_y}+\epsilon_2\frac{x^2}{R_x^2}+\epsilon_3\frac{y^2}{R_y^2}+\epsilon_4\frac{z^2}{R_z^2}\\
&+\epsilon_{5}\frac{xz}{R_{x}R_{z}
}+\epsilon_{6}\frac{yz}{R_{y}R_{z}},\\
\delta \phi=&g\left(\alpha_1\frac{xy}{R_xR_y}+\alpha_2\frac{x^2}{R_x^2}+\alpha_3\frac{y^2}{R_y^2}+\alpha_4\frac{z^2}{R_z^2}\right.\\
&\left.+\alpha_{5}\frac{xz}{R_{x}R_{z}}+\alpha_{6}\frac{yz}{R_{y}R_{z}}\right),
\end{aligned}
\label{raodongnishe}
\end{equation}
where $R_\nu=\sqrt{2\mu_0/m\omega_\nu^2}$ represents the TF radius along the $\nu$-direction ($\nu = x,\ y,\ z$). $\epsilon_i$ and $\alpha_i$ are the coefficients of corresponding excitation modes. {Substituting the perturbation ansatz Eq.~(\ref{raodongnishe}) into Eq.~(\ref{raodong}) and collecting terms with the same polynomial degree, we require that the coefficients of all independent polynomials vanish for Eq.~(\ref{raodong}) to be satisfied. This procedure results in 12 first-order differential equations for the coefficients $\epsilon_i$ and $\alpha_i$:
	\begin{widetext}
		\begin{equation}
		\label{twosetseq}
		\begin{matrix}
		\begin{aligned}
		\frac{d \epsilon_{1}}{d t}&=\omega_{x y}^{2} \alpha_{1}+2 \omega_{\eta} \epsilon_{2}-2 \omega_{\eta} \epsilon_{3}, \\
		\frac{d \epsilon_{2}}{d t}&=3 \frac{m}{m^{*}} \omega_{x}^{2} \alpha_{2}+\omega_{y}^{2} \alpha_{3}+\omega_{z}^{2} \alpha_{4}-\omega_{\eta}\epsilon_{1}, \\
		\frac{d \epsilon_{3}}{d t}&=\frac{m}{m^{*}} \omega_{x}^{2} \alpha_{2}+3 \omega_{y}^{2} \alpha_{3}+\omega_{z}^{2} \alpha_{4}+\omega_{\eta} \epsilon_{1}, \\
		\frac{d \epsilon_{4}}{d t}&=\frac{m}{m^{*}} \omega_{x}^{2} \alpha_{2}+\omega_{y}^{2} \alpha_{3}+3 \omega_{z}^{2} \alpha_{4}, \\
		\frac{d\epsilon_{5}}{d t}&=\omega_{xz}^{2}\alpha_{5}-\omega_{\eta}\epsilon_{6},\\
		\frac{d\epsilon_{6}}{d t}&=\omega_{yz}^{2}\alpha_{6}+\omega_{\eta}\epsilon_{5},\\
		\end{aligned}\ \ \ \ \  &\ \ \ \ \  \begin{aligned}
		\frac{d \alpha_{1}}{d t}&=2 \omega_{\eta} \alpha_{2}-2 \omega_{\eta} \alpha_{3}-\epsilon_{1}, \\
		\frac{d \alpha_{2}}{d t}&=-\omega_{\eta} \alpha_{1}-\epsilon_{2}, \\
		\frac{d \alpha_{3}}{d t}&=\omega_{\eta} \alpha_{1}-\epsilon_{3}, \\
		\frac{d \alpha_{4}}{d t}&=-\epsilon_{4},\\
		\frac{d\alpha_{5}}{d t}&=-\omega_{\eta}\alpha_{6}-\epsilon_{5},\\
		\frac{d\alpha_{6}}{d t}&=\omega_{\eta}\alpha_{5}-\epsilon_{6},\\
		\end{aligned} \\
		\end{matrix}
		\end{equation}
	\end{widetext}
	where we defined a new frequency $\omega_{\eta}\equiv \omega_{eff}\omega_x/\omega_y=\omega_{eff}^\prime\omega_y/\omega_x$, which characterises the coupling strength between these modes.}

\begin{figure}[t]
	\centerline{\includegraphics[width=0.45\textwidth]{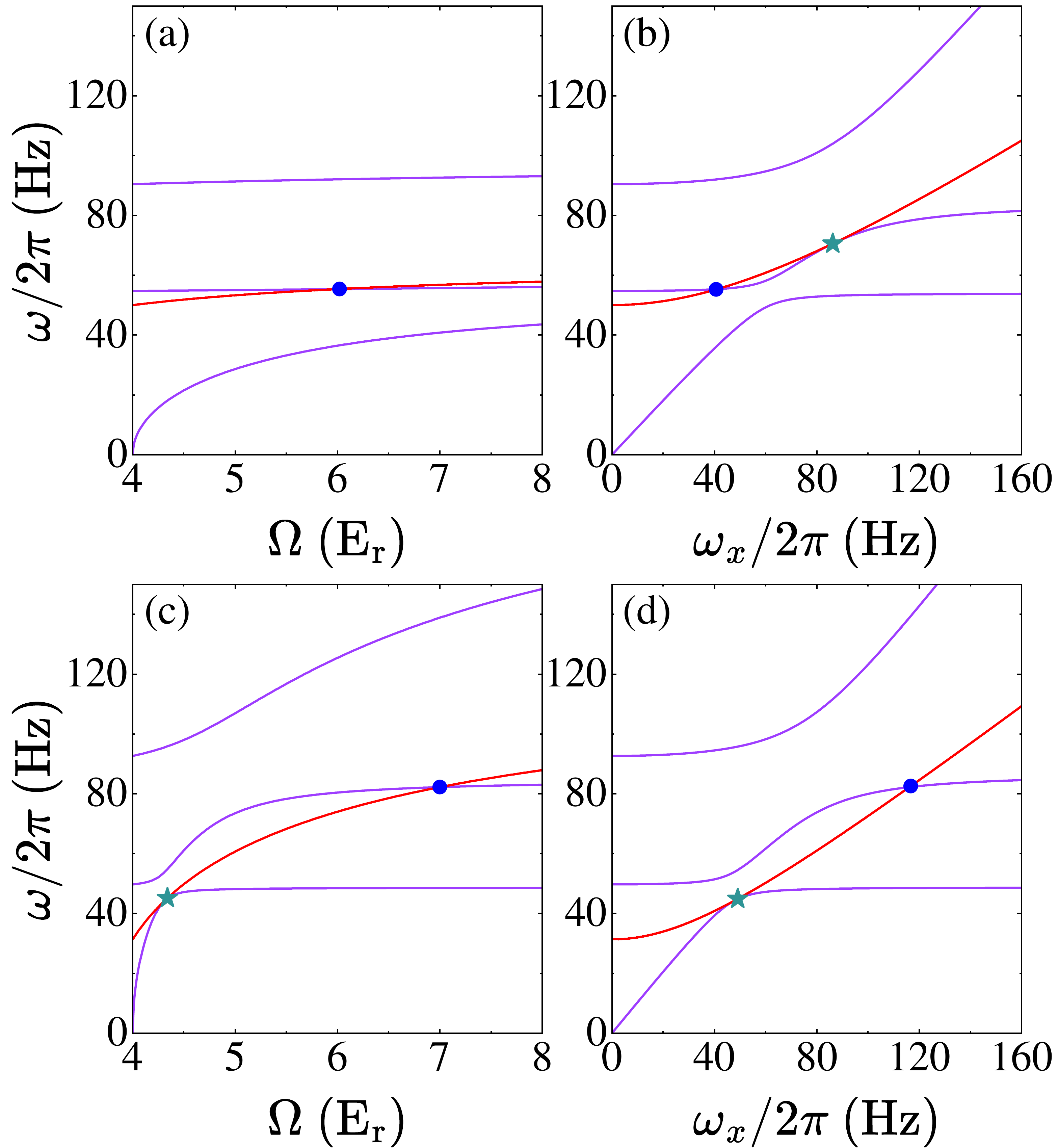}}
	\caption{The dependence of the eigenfrequencies $\omega_{1,2,3}$ (purple curves) of Eq.~(\ref{quadrupoleeigen}) and the frequency $\omega_{xy}$ (red curves) on Raman coupling strength $\Omega$ and $x$-direction trapping frequency $\omega_x$. The red line can intersect a purple line in various ways. Panel (a) shows a single intersection where $\omega_{xy}$ meets $\omega_2$ once. Panel (b) presents intersecting twice between $\omega_{xy}$ and $\omega_2$. In panels (c) and (d), $\omega_{xy}$ crosses each of $\omega_2$ and $\omega_3$ once, yielding two cross points per panel. Overall, there can be one intersection (a), two intersections (b), or intersections with two purple lines as in panels (c) and (d). In these four panels, green stars (\textcolor[RGB]{0,128,0}{\scalebox{0.8}{$\bigstar$}}) represent points satisfying the first degeneracy condition [Eq.~(\ref{degeneracy1})], while blue circles (\textcolor{blue}{$\bullet$}) represent points satisfying the second degeneracy condition [Eq.~(\ref{degeneracy2})]. For panel~(a) [(c)], the parameters are set to $(\omega_x,\omega_y,\omega_z)=(41.08,50,35)\times 2\pi$ Hz [$(116.13,14\sqrt{5},52)\times 2\pi$ Hz], while for panel~(b) [(d)], they are adjusted to $(\omega_y,\omega_z)=(50,35)\times 2\pi$ Hz [$(14\sqrt{5},52)\times 2\pi$ Hz] with $\Omega=6E_\text{r}$ ($7E_\text{r}$).}
	\label{fig1}
\end{figure}

In the absence of a synthetic magnetic field ($\eta = 0$), the three scissors modes in Eq.~(\ref{twosetseq}) are completely decoupled and the BEC exhibits independent harmonic motion in space, with the frequencies given by $\omega_{xy} = \sqrt{(m/m^*)\omega_x^2 + \omega_y^2}$, $\omega_{xz} = \sqrt{(m/m^*)\omega_x^2 + \omega_z^2}$, and $\omega_{yz} = \sqrt{\omega_y^2 + \omega_z^2}$. However, the diagonal quadrupole modes are still coupled together but no longer couple with the scissors mode. The coupling between quadrupole modes is described by the following set of equations:
\begin{equation}
\begin{aligned}
&{{\displaystyle{\frac{\partial^{2}\epsilon_{2}}{\partial t^{2}}}+3\frac{m}{m^{\ast}}{\omega_{x}^{2}}\epsilon_{2}+\omega_{y}^{2}\epsilon_{3}+\omega_{z}^{2}\epsilon_{4}=0}}, \\
&{{\displaystyle{\frac{\partial^{2}\epsilon_{3}}{\partial t^{2}}}+\frac{m}{m^{\ast}}\omega_{x}^{2}\epsilon_{2}+3\omega_{y}^{2}\epsilon_{3}+\omega_{z}^{2}\epsilon_{4}=0}},\\
&{{\displaystyle{\frac{\partial^{2}\epsilon_{4}}{\partial t^{2}}}+\frac{m}{m^{\ast}}\omega_{x}^{2}\epsilon_{2}+\omega_{y}^{2}\epsilon_{3}+3\omega_{z}^{2}\epsilon_{4}=0}}.
\end{aligned}
\label{quadrupoleeigen}
\end{equation}
Solving Eq.~(\ref{quadrupoleeigen}) yields the eigenfrequencies ($\omega_{1,2,3}$) of the three diagonal quadrupole modes. Figure~\ref{fig1} shows the dependence of these eigenfrequencies (purple curves) on  $\Omega$ and $x$-direction trapping frequency $\omega_x$. It is worth noting that in the presence of SOC and the synthetic magnetic field ($\eta \ne 0$), only the scissors mode in the $x$-$y$ plane can couple with the diagonal quadrupole modes, while the other two scissors modes $xz$ and $yz$ are not involved.

The frequency $\omega_{xy}$ represented by the red curve is also shown in Figs.~\hyperref[fig1]{\ref*{fig1}(a)}-\hyperref[fig1]{\ref*{fig1}(d)}. The crossing points between the curves are marked with green stars and blue circles to distinguish two sets of system parameters, for which the eigenfrequencies [Eq.~(\ref{quadrupoleeigen})] of the diagonal quadrupole modes and the scissors mode frequency $\omega_{xy}$ become degenerate. For the case mentioned above, this corresponds to the following two degeneracy conditions:
\begin{subequations}
	\begin{align}
	&\sqrt{\frac m{m^*}}\omega_x=\omega_y,\label{degeneracy1}\\
	\omega_z=&\sqrt{\frac25\left(\frac m{m^*}\omega_x^2+\omega_y^2\right)}.\label{degeneracy2}
	\end{align}
	\label{degeneracy}
\end{subequations}

Under the first degeneracy condition [Eq.~(\ref{degeneracy1})], the modes in the $x$-$y$ plane couple together, specifically, the scissors mode $xy$ directly interacts with the superposed mode $Q_1 = (\omega_x/\omega_y)x^2 - (\omega_y/\omega_x)y^2$. This has been studied in the previous work~\cite{PhysRevLett.120.183202}. Yet, when the second degeneracy condition [Eq.~(\ref{degeneracy2})] is satisfied, all four modes, including $xy$, $x^2$, $y^2$ and $z^2$, are coupled, and the specific manifestation of this coupling has not been investigated. We examine two scenarios corresponding to Raman coupling strength at $\Omega = 6E_\text{r}$ [Figs.~\hyperref[fig1]{\ref*{fig1}(a)} and \hyperref[fig1]{\ref*{fig1}(b)}] and $\Omega = 7E_\text{r}$ [Figs.~\hyperref[fig1]{\ref*{fig1}(c)} and \hyperref[fig1]{\ref*{fig1}(d)}], respectively. These values are far from the phase transition point $\Omega_c$ from the single-minimum phase to the plane-wave phase, also making them convenient for GPE simulation. Under these conditions, we observe that there is only one crossing point in Fig.~\hyperref[fig1]{\ref*{fig1}(a)}, indicating that only the second degeneracy can occur. In Figs.~\hyperref[fig1]{\ref*{fig1}(a)}-\hyperref[fig1]{\ref*{fig1}(d)}, crossing points correspond to the first and the second degeneracy conditions. We next analyze the second degeneracy case.

\subsection{Coupling between $xy$, $x^2$, $y^2$ and $z^2$}

\begin{figure}[t]
	\centerline{
		\includegraphics[width=0.5\textwidth]{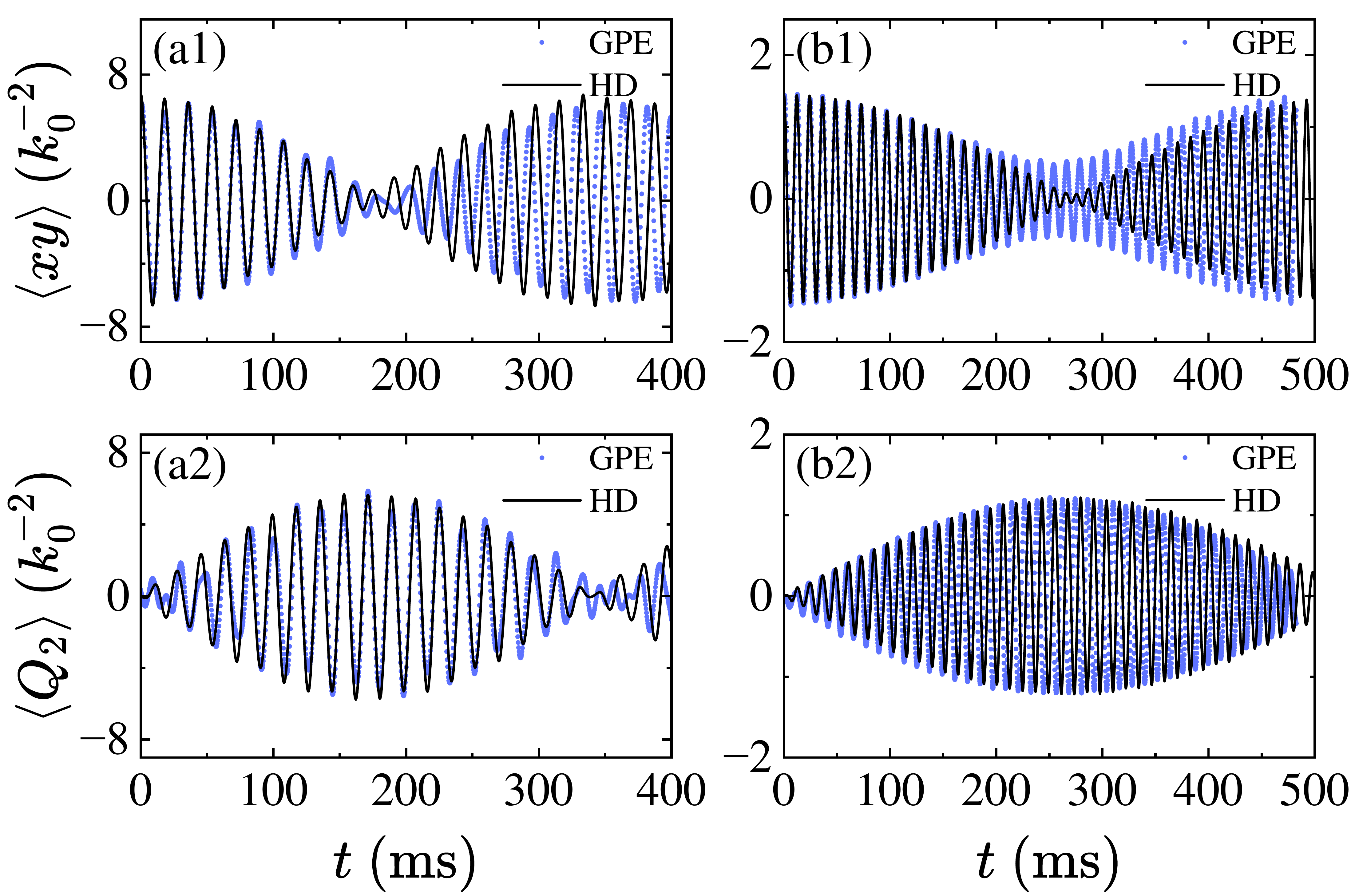}}
	\caption{Time evolution of the scissors mode $\langle xy \rangle$ and the superposed quadrupole mode $\langle Q_2\rangle$ obtained from GPE numerical simulation (dots) and HD theory (solid lines) in the presence of {a detuning gradient} $\eta=0.001E_\text{r}$. For (a1-a2) [(b1-b2)], the Raman coupling strength is $\Omega=6E_\text{r}$ ($7E_\text{r}$), and the trapping frequency is set as $(\omega_x, \omega_y, \omega_z) = (41.08, 50, 35)\times 2\pi$ Hz [$(116.13, 14\sqrt{5}, 52)\times 2\pi$ Hz]. The dynamics in panel (a) [panel (b)] correspond to the blue circles in Figs.~\hyperref[fig1]{\ref*{fig1}(a)} and \hyperref[fig1]{\ref*{fig1}(b)} [Figs.~\hyperref[fig1]{\ref*{fig1}(c)} and \hyperref[fig1]{\ref*{fig1}(d)}]. {In our simulation, the parameter of the particle number $N = 2\times 10^4$ was used, and this parameter remained unchanged in the subsequent simulations.} The system's dynamics are initiated by an abrupt rotation on the harmonic trap with a slight angle $\varphi_0=3^\circ$ in the $x$-$y$ plane. }
	\label{fig2}
\end{figure}

Bringing the second degeneracy condition [Eq.~(\ref{degeneracy2})] into Eq.~(\ref{quadrupoleeigen}), one of the eigenfrequencies is equal to the scissors mode frequency $\omega_{xy}$ in magnitude (blue cricles in Fig.~\ref{fig1}). This result indicates that there is a mode referred to as $Q_2$, expressed as a superposition of three coefficients $\epsilon_{1,2,3}$:
\begin{equation}
\langle Q_2\rangle = -\frac{1}{5}\frac{\frac{m}{m^*}\omega_x^2+\omega_y^2}{\frac{m}{m^*}\omega_x^2-\omega_y^2}(\epsilon_2-\epsilon_3)+\epsilon_4.
\label{Q21}
\end{equation}
Using the ansatz $\delta n$ from Eq.~(\ref{raodongnishe}), solving inversely for the three coefficients {$\epsilon_{2,3,4}$} yields their forms in terms of the three measurable quantities $x^2$, $y^2$ and $z^2$. The relationships between the coefficients and the measurable quantities are as follows:
\begin{equation}
\begin{aligned}
\epsilon_2&=\frac{2}{5}\frac{\omega_x}{\omega_y}\langle x^2\rangle - \frac{1}{10}\frac{\omega_y}{\omega_x}\langle y^2\rangle - \frac{1}{10}\frac{\omega_z^2}{\omega_x\omega_y}\langle y^2\rangle,\\
\epsilon_3&=-\frac{1}{10}\frac{\omega_x}{\omega_y}\langle x^2\rangle + \frac{2}{5}\frac{\omega_y}{\omega_x}\langle y^2\rangle - \frac{1}{10}\frac{\omega_z^2}{\omega_x\omega_y}\langle y^2\rangle,\\
\epsilon_4&=-\frac{1}{10}\frac{\omega_x}{\omega_y}\langle x^2\rangle - \frac{1}{10}\frac{\omega_y}{\omega_x}\langle y^2\rangle + \frac{2}{5} \frac{\omega_z^2}{\omega_x\omega_y}\langle y^2\rangle,\\
\end{aligned}
\label{xishu}
\end{equation}
where $\langle \cdot\rangle$ represents an average with respect to the density fluctuations $\delta n$. Substituting Eq.~(\ref{xishu}) into Eq.~(\ref{Q21}), \textcolor{red}{w}e can obtain the operator form corresponding to the $Q_2$ mode as follows:
\begin{equation}
Q_2=\frac{(m \omega_x^3)/(m^*\omega_y) x^2-(\omega_y^3/\omega_x) y^2}{5(\omega_y^2-m\omega_x^2/m^*)}  + \frac{4\omega_{xy}^2}{25\omega_x\omega_y} z^2.
\end{equation}
Following the same procedure, the average of the scissors mode $xy$ is expressed as
\begin{equation}
\langle xy \rangle = \frac{4}{105} \pi R_x^2 R_y^2 R_z \epsilon_1.
\end{equation}

Introducing a weak synthetic magnetic field ($\eta = 0.001E_\text{r}$) couples two independent modes, $Q_2$ and $xy$, breaking the second degeneracy condition. {By rotating the BEC by a small angle around the $z$-axis, the $xy$ mode is excited. The subsequent dynamics of the two modes ($xy$ and $Q_2$) exhibits a characteristic beating effect.} Figure~\ref{fig2} shows the coupling between the scissors mode $xy$ and the superposed mode $Q_2$. The black lines and blue dots, which are the results of HD theory and GPE simulation respectively, are in good agreement. The black lines and blue dots, which are the results of HD theory and GPE simulation respectively, are in good agreement. {The discrepancy observed in Fig.~\hyperref[fig2]{\ref*{fig2}(b1)} near the beat node arises from the finite particle number. We have verified that as the particle number N increases, the system approaches the TF approximation more closely, and these mismatches are significantly reduced.} Comparing Figs.~\hyperref[fig2]{\ref*{fig2}(a)} and \hyperref[fig2]{\ref*{fig2}(b)}, the beat frequency when $\Omega = 6E_\text{r}$ is larger than that when $\Omega = 7E_\text{r}$, indicating stronger mode coupling in the former case. This is because increasing the Raman coupling strength $\Omega$ weakens the effective synthetic magnetic field.

\subsection{Coupling between scissors modes $xz$ and $yz$}

In the previous subsection, we explored the coupling among four quadrupole modes. Now, we turn our attention to the remaining two scissors modes, $xz$ and $yz$, which also couple under the second degeneracy condition. In the absence of a synthetic magnetic field ($\eta=0$), the first degeneracy condition [Eq.~(\ref{degeneracy1})] implies $\omega_{xz} = \omega_{yz}$. However, under the second degeneracy condition [Eq.~(\ref{degeneracy2})], $\omega_{xz} \neq \omega_{yz}$, no degeneracy occurs between these two modes. Introducing a {weak} synthetic magnetic field ($\eta = 0.001 E_\text{r}$) will cause these two modes coupling together regardless of whether their frequencies are equal or not. The response of the system differs depending on whether the frequencies are degenerate. In the following, we will analyze the case where $\omega_{xz} \neq \omega_{yz}$, and these two scissors modes are non-degenerate.

{R}otating the BEC around the $y$-axis excites the $xz$ mode, while rotating around the $x$-axis excites the $yz$ mode. From Eq.~(\ref{twosetseq}), we derive the coupling equations for these two modes as follows:
\begin{equation}
\begin{aligned}
&\frac{d^{2}\epsilon_{5}}{dt^{2}}+\omega_{xz}^{2}\left(1-\frac{\omega_{\eta}^2}{\omega_{yz}^{2}}\right)\epsilon_{5}+\omega_{\eta}\left(1+\frac{\omega_{xz}^{2}}{\omega_{yz}^{2}}\right)\frac{d\epsilon_{6}}{d t}=0,\\
&\frac{d^{2}\epsilon_{6}}{d t^{2}}+\omega_{yz}^{2}\left(1-\frac{\omega_{\eta}^2}{\omega_{xz}^{2}}\right)\epsilon_{6}-\omega_{\eta}\left(1+\frac{\omega_{yz}^{2}}{\omega_{xz}^{2}}\right)\frac{d\epsilon_{5}}{d t}=0,
\end{aligned}
\label{xzyzeq}
\end{equation}
where the coefficients are determined by $\omega_{xz}$, $\omega_{yz}$, and $\omega_\eta$. It is worth noting that if $\omega_{xz} = \omega_{yz}$, the coefficients for $\epsilon_5$ and $\epsilon_6$ become identical, satisfying the first degeneracy condition. In this degenerate case, the motion is linked to a BEC gyroscope as discussed in Ref.~\cite{PhysRevA.108.053316}. Additionally, similar coupling structures can also be observed in the dipole modes of the system, where the motion resembles a Foucault pendulum in the $x$-$y$ plane, as reported in Ref.~\cite{PhysRevLett.120.183202}. When $\omega_{xz} \neq \omega_{yz}$, the trajectories of the two modes form various Lissajous-like patterns in the $\langle xz \rangle$-$\langle yz \rangle$ parameter space.

\begin{figure}[t]
	\centerline{
		\includegraphics[width=0.48\textwidth]{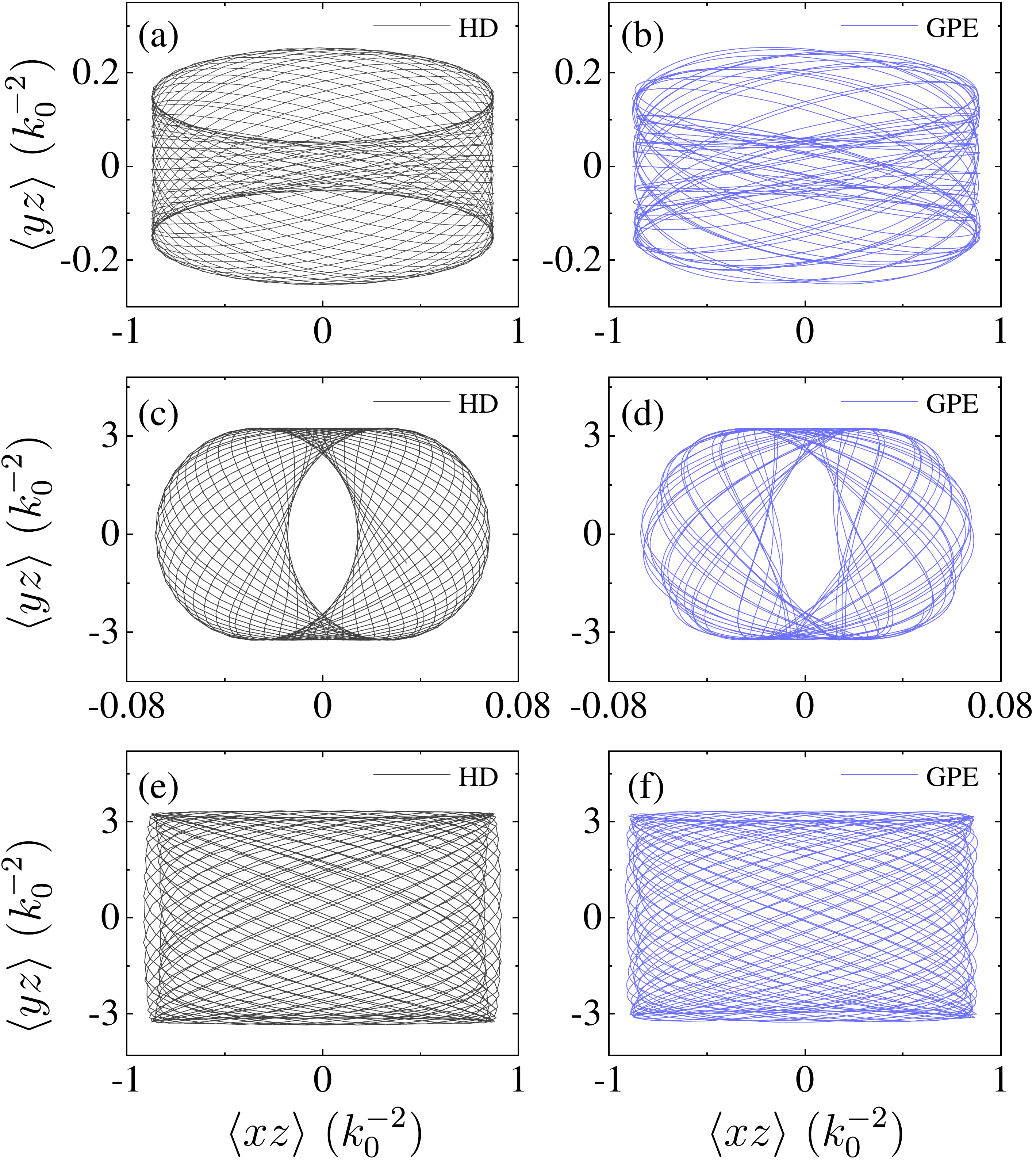}}
	\caption{Illustration of the Lissajous-like patterns obtained from HD theory (left) and GPE simulations (right). The system parameters are chosen as $(\omega_x, \omega_y, \omega_z)=(116.13, 14\sqrt{5}, 52)\times 2\pi$ Hz, and $\Omega=7E_\text{r}$, fulfilling the condition indicated by blue circles \textcolor{blue}{$\bullet$} in Figs.~\hyperref[fig1]{\ref*{fig1}(c)} and \hyperref[fig1]{\ref*{fig1}(d)}. Panels~(a), (c) and (d) depict the parameters with initial set $(\epsilon_5,\epsilon_6)=(1,0),\ (0,1)$ and $(1,1)$, while panels~(b), (d) and (f) show the GPE simulation. In the GPE simulation, panel~(b) depicts an abrupt $3^\circ$ rotation in the $x$-$z$ plane, induced by rotating the harmonic trap. Panel~(d) shows a similar $3^\circ$ rotation in the $y$-$z$ plane. Panel~(f) illustrates a superposed rotation of $3^\circ$ in both the $x$-$z$ and $y$-$z$ planes.}
	\label{fig3}
\end{figure}

We adopt three excitation methods to generate these Lissajous-like patterns: (i) Rotation around the $y$-axis, corresponding to $\epsilon_5 = 1$, $\epsilon_6 = 0$ in the HD theory. (ii) Rotation around the $x$-axis, corresponding to $\epsilon_5 = 0$, $\epsilon_6 = 1$. (iii) Successive rotations around the $x$- and $y$-axes, corresponding to $\epsilon_5 = 1$ and $\epsilon_6 = 1$. Figure~\ref{fig3} shows the resulting Lissajous-like patterns generated by these three methods, with the HD theory and GPE simulations showing good agreement in their spatial paths.

{Under the second degeneracy condition [Eq.~(\ref{degeneracy2})], the two scissors modes $ xz $ and $ yz $ are generally not degenerate, except when $\sqrt{m/m^*}\omega_x=\omega_y$. As shown in Eq.~\eqref{xzyzeq}, the detuning gradient $\eta$ induces a coupling between these two modes. However, due to the significant frequency gap between them, the coupling remains weak, resulting in only small perturbations to their independent dynamics. This weak coupling leads to distinct Lissajous-like patterns, which depend sensitively on the initial conditions. The dominant mode in the dynamics can be identified from both the amplitude of the time evolutionand the weights of the peaks in the Fourier spectrum. For instance, when the $\langle xz \rangle$ mode is initially excited with a large amplitude, the Fourier spectrum is dominated by the frequency component $ \omega_{xz} $ with a secondary, weaker peak near $\omega_{yz}$ [see Fig.~\hyperref[fig3]{\ref*{fig3}(a)-(b)}]. Conversely, when the $\langle yz \rangle$ mode is primarily excited, the Fourier spectrum exhibits a dominant frequency component at $ \omega_{yz} $, accompanied by a smaller peak appears near $\omega_{xz}$ [see Fig.~\hyperref[fig3]{\ref*{fig3}(c)-(d)}]. When both modes are equally excited, two prominent peaks with similar weigths appear in the Fourier spectrum [Fig.~\hyperref[fig3]{\ref*{fig3}(e)-(f)}]. The substantial frequency difference between the two modes modifies the coefficients in the coupling equations, thereby preventing the emergence of a perfect beating effect~\cite{RevModPhys.68.755}. Instead, the weak interaction between the two modes gives rise to distinct Lissajous-like patterns, highlighting the interplay between the two modes and the role of detuning-gradient-induced coupling.}

\begin{figure}[t]
	\centerline{
		\includegraphics[width=0.46\textwidth]{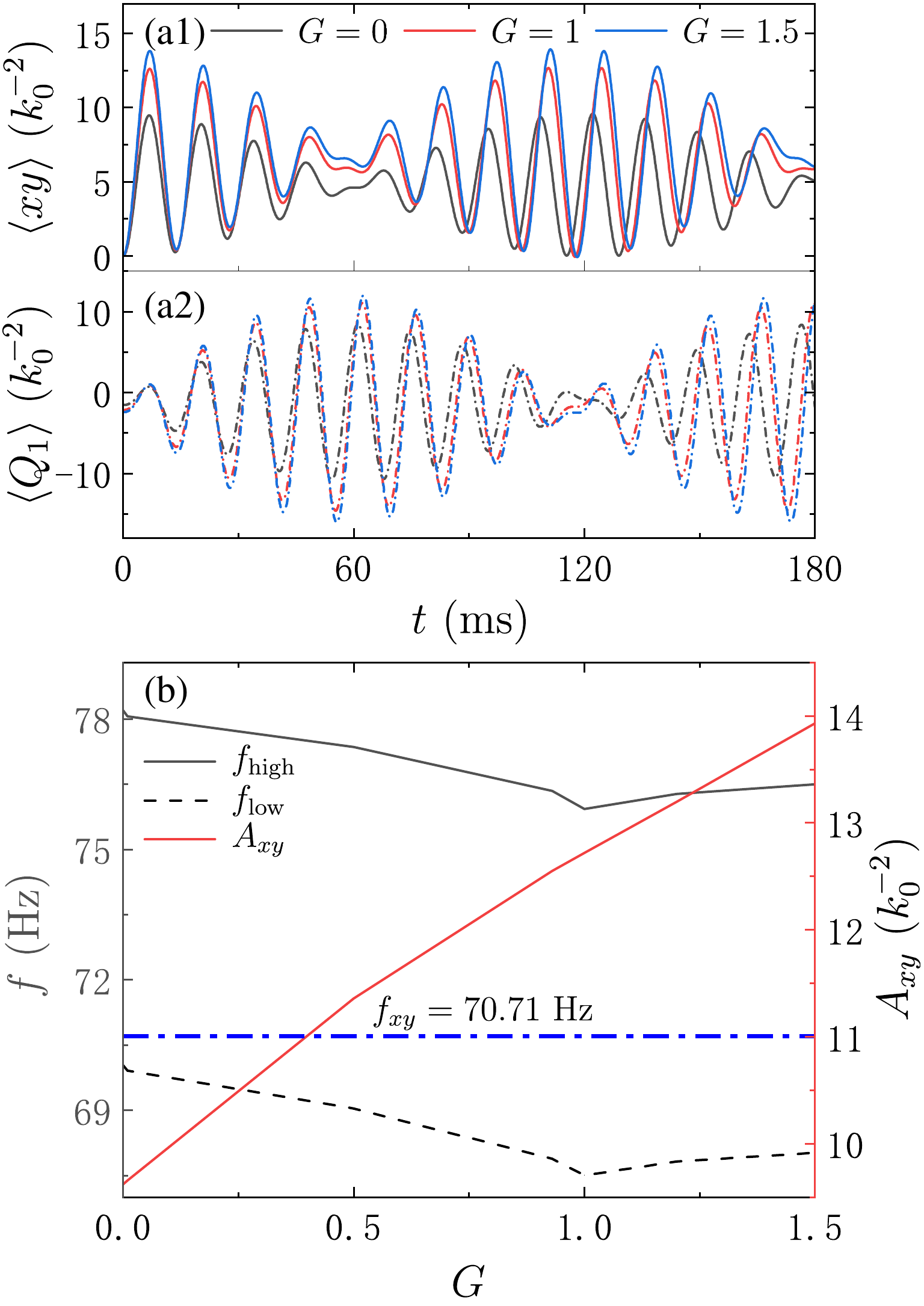}}
	\caption{Panels~(a1-a2) shows three sets of beat patterns of the scissors mode $xy$ and the superposed mode $Q_1$ under different interaction parameters $(G = 0, 1, 1.5)$, displayed in different colors. The system parameters are chosen as $ (\omega_x, \omega_y, \omega_z) = (50\sqrt{3}, 50, 35)\times 2\pi $ Hz and $\Omega = 6E_\text{r}$, ensuring the first degeneracy condition [Eq.~(\ref{degeneracy1})] is met.  In panel~(b), the black solid line and black dashed line represent the positions of two main peaks in the Fourier spectrum of the beat pattern of the scissors mode $ xy $ as the interaction ($G$) varies. The blue dashed line indicates the scissors mode frequency ($f_{xy} = 70.71$ Hz) obtained using HD theory in the absence of synthetic magnetic field ($\eta=0 $) and $ G = 1 $. The red solid line represents {the envelope function amplitude} $A_{xy}$ of the scissors mode $ xy $ as a function of $G$, showing a linear relationship. The dynamics are initiated by abruptly rotating the harmonic trap by a small angle $\phi_0 = 3^\circ$ in the $x$-$y$ plane.}
	\label{fig4}
\end{figure}

\section{Effect of the anisotropic interaction}\label{sec4}

In this section, we explore how anisotropic interactions affect the coupling between quadrupole modes, specifically under the first degeneracy condition [Eq.~(\ref{degeneracy1})]. The synthetic magnetic field breaks the initial degeneracy, coupling the $xy$ and $Q_1$ modes, which leads to a beating effect. Changes in the inter-species interaction $g_{12}$ directly impact the mode frequencies and amplitudes, reflecting shifts in the system's energy distribution.

\begin{figure}[t]
	\centerline{
		\includegraphics[width=0.35\textwidth]{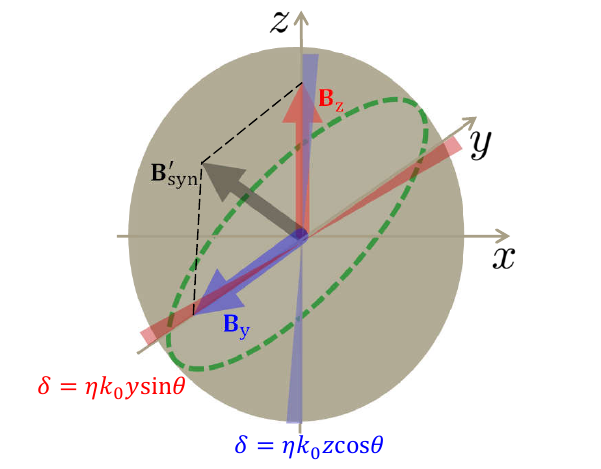}}
	\caption{Illustration of the synthetic magnetic field for a BEC with the detuning dependent on both the $y$ and $z$ positions. Specifically, $y$ ($z$)-position dependent detuning generate a $+z$ ($-y$)-direction synthetic magnetic field $\textbf{B}_{z}$ $
		(\textbf{B}_{y})$ represented by red (blue) arrow . The black arrow, $\textbf{B}_{\text{syn}}^\prime$, denotes the total synthetic magnetic field resulting from the superposition of $\textbf{B}_y$ and $\textbf{B}_z$. The green dashed line marks the plane perpendicular to $\textbf{B}_{\text{syn}}^\prime$.}
	\label{fig5}
\end{figure}

For simplicity, we initially assume isotropic interactions with $g_{ij} = g$. However, the present SOC BEC experiment allows the scattering length to be tuned by the Feshbach resonance, allowing the exploration of various interaction parameters~\cite{PhysRevLett.85.1795,PhysRevLett.107.073202}. In different parameter regions, SOC BEC exhibits new properties. For example, supersolid phase will appear in antiferromagnetic configuration~\cite{PhysRevLett.108.225301,li2017stripe}. The phase separation and miscibility induced by the interaction make the ground state be a necklace state and a continuous flow state, respectively~\cite{PhysRevA.95.041604}. To rigorously examine the influence of anisotropic interactions, we focus on the first degeneracy condition [Eq.~(\ref{degeneracy1})] and investigate how varying interaction parameter $G = g_{12}/g$ influences quadrupole mode coupling. The system parameters are set to $(\omega_x, \omega_y, \omega_z) = (50\sqrt{3}, 50, 35)\times 2\pi$ Hz and $\Omega = 6E_\text{r}$, where the scissors mode $xy$ couples with the diagonal quadrupole mode $Q_1 = (\omega_x/\omega_y)x^2 - (\omega_y/\omega_x)y^2$. Numerical simulations reveal the time evolution of both modes.

\begin{figure*}[t]
	\centerline{
		\includegraphics[width=0.9\textwidth]{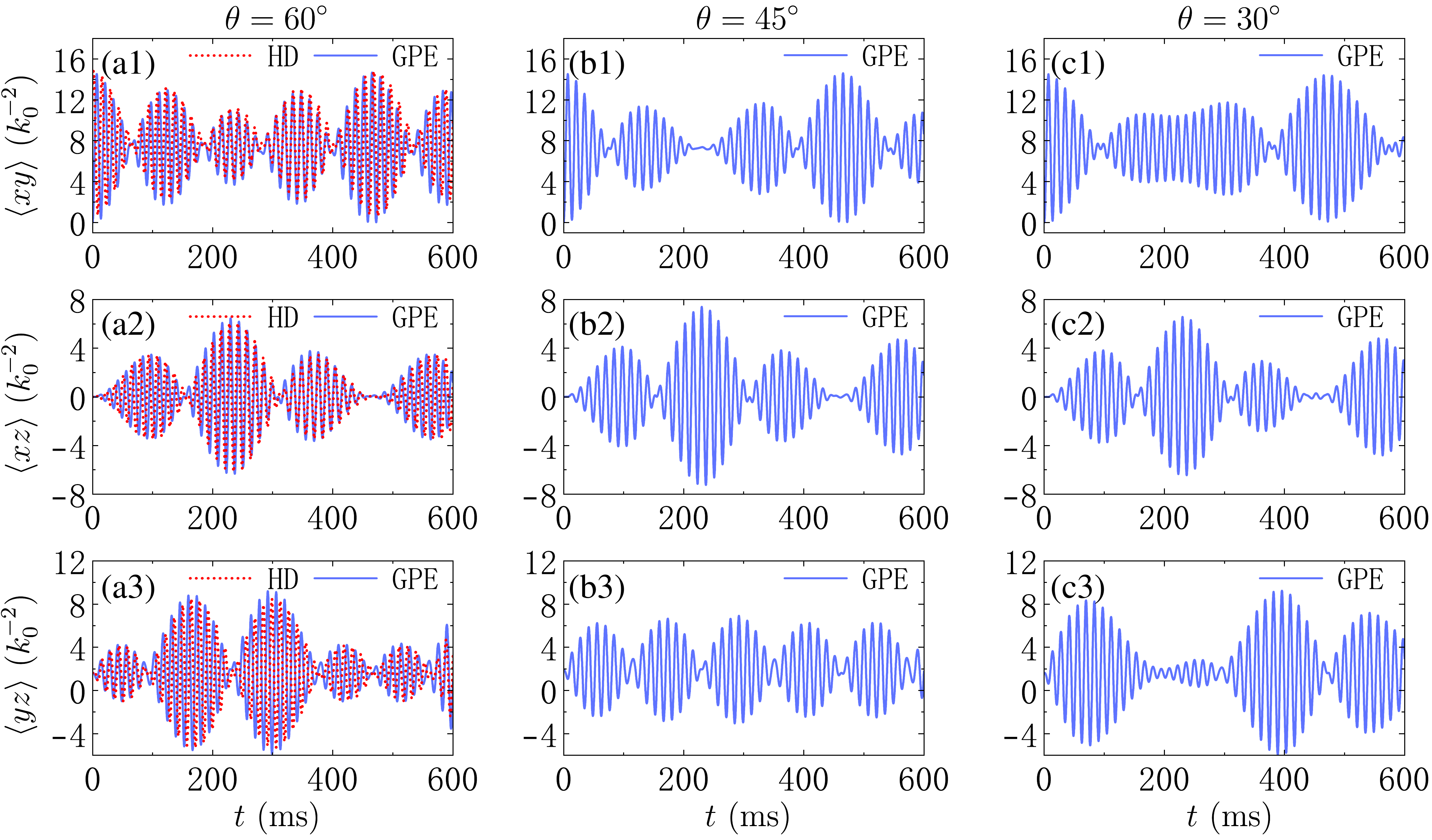}}
	\caption{Schematic representation of the scissors modes $\langle xy\rangle$, $\langle xz\rangle$, $\langle yz\rangle$ under a synthetic magnetic field applied in the $y$-$z$ plane with three different gradient ratios, corresponding to $\theta = 60^\circ$ (left), $\theta = 45^\circ$ (middle) and $\theta = 30^\circ$ (right), respectively. The parameters for the GPE (blue line) and HD (red dotted line) simulations are: $(\omega_x, \omega_y, \omega_z) = (50\sqrt{3}, 50, 50)\times 2\pi$ Hz, $\Omega = 6E_\text{r}$, and $\eta = 0.001E_\text{r}$. The system's dynamics are triggered by a sudden $\varphi_0=3^\circ$ rotation of the harmonic trap in the $x$-$y$ plane.
	}
	\label{fig6}
\end{figure*}

Figure~\hyperref[fig4]{\ref*{fig4}(a)} shows the time evolution of the $xy$ and $Q_1$ modes for $G = 0, 1$, and $1.5$. These modes remain coupled, indicating that the beating effect persists even when $G \neq 1$. The variation in the inter-species interaction $g_{12}$ alters the size of the atomic cloud, causing a linear change in {the envelope function amplitude} $A_{xy}$ of the scissors mode, as shown by the red solid line in Fig.~\hyperref[fig4]{\ref*{fig4}(b)}. Additionally, changes in $g_{12}$ shift the oscillation frequencies of the quadrupole modes. To highlight these frequency shifts, we performed a Fourier transform on the time-evolution data {of the $xy$ mode}. In Fig.~\hyperref[fig4]{\ref*{fig4}(b)}, when $\eta \neq 0$, the degeneracy is lifted, resulting in two distinct frequencies near the scissors mode frequency $f_{xy} = 70.71$ Hz (blue dashed line). The beating effect arises from the frequency difference between these two modes. We further plot the frequencies of the two dominant peaks in the spectrum, $f_{\text{high}}$ (black solid line) and $f_{\text{low}}$ (black dashed line), as functions of the interaction parameter. Both frequencies exhibit a minimum at $G = 1$.

\section{Effect of Synthetic Magnetic Field Direction}\label{sec5}


In previous sections, we examined the quadrupole mode under the influence of a synthetic magnetic field {$\textbf{B}_{\text{syn}}$}, which originates from a detuning gradient dependent solely on the $y$-position. However, the detuning's position dependence can be generalized {[see Eq.~\eqref{SynB}]}. In this section, we introduce a $z$-position-dependent term alongside the $y$-position-dependent term in the detuning. This generates a new synthetic magnetic field along the $-y$ direction. As illustrated in Fig.~\ref{fig5}, the black arrows represent the total synthetic magnetic field {$\textbf{B}_{\text{syn}}^\prime$} induced by the modified position-dependent detuning, while the red (blue) arrows indicate the contributions from the $y$ ($z$)-position-dependent terms. In this scenario, the last term in Eq.~(\ref{eq2}) is modified as follows:
\begin{equation}
-\eta k_0 y\sigma_z\to-\eta k_0\left(y\sin\theta +z\cos\theta \right)\sigma_z,
\end{equation}
where $\eta$ is the coefficient of the position-dependent detuning, and $\theta$ is a dimensionless parameter that tunes the ratio between the $y$ and $z$ detuning components. By varying $\theta$, we can explore how different directions of the synthetic magnetic field affect the collective mode dynamics, as shown in Eq.~\eqref{twosetseq2}.

{When the parameter $\theta = 0^\circ$, the detuning becomes $z$-dependent and the synthetic magnetic field points to the $y$-direction. Under the first degeneracy condition [Eq.(\ref{degeneracy1})], the coupling dynamics of the three scissors modes are similar to the conclusion in Sec.~\ref{sec3}: the $xz$ scissors mode becomes coupled to another superposed-mode $Q_1^\prime = (\omega_x/\omega_z)x^2 - (\omega_z/\omega_x)z^2$, and the scissors modes $xy$ and $yz$ are coupled together, forming a characteristic beating effect.
	Figure~\ref{fig6} illustrates the time evolution for three scissors modes under different angles $\theta$. Specifically, for $\theta = 60^\circ$ (left), $45^\circ$ (middle), and $30^\circ$ (right), distinct differences in the beating patterns, both in oscillation periods and shapes, are observed. Notably, the case of $\theta=60^\circ$ Fig.~\hyperref[fig6]{\ref*{fig6}(a1-a3)} shows excellent agreement between the generalized HD theory (see Appendix) and GPE simulations.} Even minor alterations in the gradient direction lead to pronounced beating effects, underlining the high sensitivity of the system to the direction of the synthetic magnetic field. This unique property empowers the SOC BEC system not only to detect the existence of {a magnetic field gradient} but also to precisely ascertain its direction, thus making it a promising candidate for quantum sensing applications.

\section{Conclusions}\label{sec6}

In summary, we have conducted a systematic study of collective oscillations of SOC BEC under the position-dependent detuning. By combining HD theory with numerical simulations of the GPE, we investigated the coupling dynamics of quadrupole modes with various degeneracy conditions. We identified a superposition of diagonal quadrupole modes $Q_2$ coupling with the scissors mode $xy$, resulting in interesting beating effects. In contrast, the coupling of the scissors modes $xz$ and $yz$ forms trajectories reminiscent of Lissajous patterns. Additionally, we extended our studies to more general scenarios and numerically investigated the effects of anisotropic interactions and the direction of the synthetic magnetic field on the dynamics of quadrupole modes. Our results indicate that anisotropic interactions significantly affect the amplitudes and oscillation frequencies of the scissors modes. Varying the direction of the synthetic magnetic field induces a notable response from the system, with significant alterations in the beating dynamics of the scissors modes. This suggests that the system has potential as a magnetic field gradiometer for measuring the direction of {magnetic field gradients}. These findings not only provide new insights into the fundamental physics of SOC BEC but also offer valuable theoretical foundations for future experimental research and application development.

\begin{acknowledgments}
	This work is supported by the National Natural Science Foundation of China (NSFC) under Grants No. 12374247 and No. 11974235, as well as by the Shanghai Municipal Science and Technology Major Project (Grant No. 2019SHZDZX01-ZX04). C.Q. is supported by ACC-New Jersey under Contract No. W15QKN-18-D-0040.
	\\
	
\end{acknowledgments}


\appendix
{
	\section{Hydrodynamic equations for synthetic magnetic field in arbitrary directions}
	
	By applying the perturbation ansatz in Eq.~\eqref{raodongnishe} (Sec.~\ref{sec5}), we derive 12 revised coupled equations governing the six collective modes. These equations include additional terms involving $\omega_{\eta_2}$, which originate from the influence of the magnetic field pointing to an arbitrary direction in the $y-z$ plane. The complete set of equations is presented below:
	\begin{widetext}
		\begin{equation}
		\label{twosetseq2}
		\begin{matrix}
		\begin{aligned}
		\frac{d \epsilon_{1}}{d t}&=\omega_{x y}^{2} \alpha_{1}+2 \omega_{\eta_1} \epsilon_{2}-2 \omega_{\eta_1} \epsilon_{3}{-\omega_{\eta_2}\epsilon_6}, \\
		\frac{d \epsilon_{2}}{d t}&=3 \frac{m}{m^{*}} \omega_{x}^{2} \alpha_{2}+\omega_{y}^{2} \alpha_{3}+\omega_{z}^{2} \alpha_{4}-\omega_{\eta_1}\epsilon_{1}{-\omega_{\eta_2}\epsilon_5}, \\
		\frac{d \epsilon_{3}}{d t}&=\frac{m}{m^{*}} \omega_{x}^{2} \alpha_{2}+3 \omega_{y}^{2} \alpha_{3}+\omega_{z}^{2} \alpha_{4}+\omega_{\eta} \epsilon_{1}, \\
		\frac{d \epsilon_{4}}{d t}&=\frac{m}{m^{*}} \omega_{x}^{2} \alpha_{2}+\omega_{y}^{2} \alpha_{3}+3 \omega_{z}^{2} \alpha_{4}{+\omega_{\eta_2}\epsilon_5}, \\
		\frac{d\epsilon_{5}}{d t}&=\omega_{xz}^{2}\alpha_{5}-\omega_{\eta_1}\epsilon_{6}{+2\omega_{\eta_2}\epsilon_2-2\omega_{\eta_2}\epsilon_4},\\
		\frac{d\epsilon_{6}}{d t}&=\omega_{yz}^{2}\alpha_{6}+\omega_{\eta_1}\epsilon_{5}{+\omega_{\eta_2}\epsilon_1},\\
		\end{aligned}\  &\  \begin{aligned}
		\frac{d \alpha_{1}}{d t}&=2 \omega_{\eta_1} \alpha_{2}-2 \omega_{\eta_1} \alpha_{3}{-\omega_{\eta_2}\alpha_6}-\epsilon_{1}, \\
		\frac{d \alpha_{2}}{d t}&=-\omega_{\eta_1} \alpha_{1}{-\omega_{\eta_2}\alpha_5}-\epsilon_{2}, \\
		\frac{d \alpha_{3}}{d t}&=\omega_{\eta_1} \alpha_{1}-\epsilon_{3}, \\
		\frac{d \alpha_{4}}{d t}&={\omega_{\eta_2}\alpha_5}-\epsilon_{4},\\
		\frac{d\alpha_{5}}{d t}&=-\omega_{\eta_1}\alpha_{6}{+\omega_{\eta_2}\alpha_2-\omega_{\eta_2}\alpha_4}-\epsilon_{5},\\
		\frac{d\alpha_{6}}{d t}&=\omega_{\eta_1}\alpha_{5}{+\omega_{\eta_2}\alpha_1}-\epsilon_{6}.\\
		\end{aligned} \\
		\end{matrix}
		\end{equation}
	\end{widetext}
	We have also introduced new representations to express these coupling coefficients:
	\begin{equation}
	\left\{
	\begin{aligned}
	\omega_{\eta_1}&=\frac{\Omega_{c}\eta}{\hbar\Omega}\frac{\omega_{x}\omega_{y}}{\omega_{xy}^{2}}\sin\theta,\\
	\omega_{\eta_2}&=\frac{\Omega_{c}\eta}{\hbar\Omega}\frac{\omega_{x}\omega_{z}}{\omega_{xz}^{2}}\cos\theta,\\
	\end{aligned}
	\right.
	\end{equation}
	where $\omega_{\eta_1}$ and $\omega_{\eta_2}$ are newly defined frequencies that characterise the coupling strength between the respective modes.
}
\nocite{*}

\bibliography{ref.bib} 

\end{document}